\documentclass[aps,prb,twocolumn,superscriptaddress,floatfix]{revtex4}
\usepackage{graphicx}
\usepackage{amsmath}
\usepackage{amsfonts}
\usepackage{amssymb}

\newcommand{\be}{\begin{equation}}
\newcommand{\ee}{\end{equation}}

\newcommand{\omegaWS}{\ensuremath{\omega_{\mathrm{B}}}}

\input{tcilatex}

\begin{document}

\title{Voltage-Current curves for small Josephson junction arrays}
\author{B. Dou\c{c}ot}
\affiliation{Laboratoire de Physique Th\'{e}orique et Hautes \'Energies, CNRS UMR 7589,
Universit\'{e}s Paris 6 et 7, 4, place Jussieu, 75252 Paris Cedex 05 France}
\author{L.B. Ioffe}
\affiliation{Center for Materials Theory, Department of Physics and Astronomy, Rutgers
University 136 Frelinghuysen Rd, Piscataway NJ 08854 USA}

\begin{abstract}
We compute the current voltage characteristic of a chain of identical
Josephson circuits characterized by a large ratio of Josephson to charging
energy that are envisioned as the implementation of topologically protected
qubits. We show that in the limit of small coupling to the environment it
exhibits a non-monotonous behavior with a maximum voltage followed by a
parametrically large region where $V\propto 1/I$. ~We argue that its
experimental measurement provides a direct probe of the amplitude of the
quantum transitions in constituting Josephson circuits and thus allows their
full characterization.
\end{abstract}

\maketitle

\section{Introduction}

In the past years, the dramatic experimental progress in the design and
fabrication of quantum two level systems in various superconducting circuits~%
\cite{Devoret04} has raised a hope that such solid state devices could
eventually serve as basic logical units in a quantum computer (qubits).
However, a very serious obstacle on this path is the ubiquitous decoherence,
which in practice limits the typical life-time of quantum superpositions of
two distinct logical states of a qubit to microseconds. This is far from
being sufficient to satisfy the requirements for implementing quantum
algorithms and providing systematic error correction.~\cite{Knill05}

This has motivated us to propose some alternative ways to design Solid-State
qubits, that would be much less sensitive to decoherence than those
presently available. These protected qubits are finite size Josephson
junction arrays in which interactions induce a degenerate ground-state space
characterized by the remarkable property that all the local operators
induced by couplings to the environment act in the same way as the identity
operator. These models fall in two classes. The first class is directly
inspired by Kitaev's program of topological quantum computation,~\cite%
{Kitaev03} and amounts to simulating lattice gauge theories with small
finite gauge groups by a large Josephson junction lattice.~\cite%
{Ioffe02,Doucot03a,Doucot03b} The second class is composed of smaller arrays
with sufficiently large and non-Abelian symmetry groups allowing for a
persistent ground-state degeneracy even in the presence of a noisy
environment.~\cite{Feigelman04,Doucot05} All these systems share the
property that in the classical limit for the local superconducting phase
variables (i.e. when the Josephson coupling is much larger than the charging
energy), the ground-state is highly degenerate. The residual quantum
processes within this low energy subspace lift the classical degeneracy in
favor of macroscopic coherent superpositions of classical ground-states. The
simplest example of such system is based on chains of rhombi (Fig. \ref%
{circuitry}) frustrated by magnetic field flux $\Phi =\Phi _{0}/2$ that
ensures that in the classical limit each rhombus has two degenerate states.~%
\cite{Doucot05}

Practically, it is important to be able to test these arrays and optimize
their parameters in relatively simple experiments. In particular one needs
the means to verify the degeneracy of the classical ground states, the
presence of the quantum coherent processes between them and measure their
amplitude. Another important parameter is the effective superconducting
stiffness of the fluctuating rhombi chain. The classical degeneracy and
chain stiffness can be probed by the experiments discussed in \cite%
{Protopopov04}; they are currently being performed~\cite{Pannetier2007}. The
idea is that a chain of rhombi threaded individually by half a
superconducting flux quantum, the non-dissipative current is carried by
charge $4e$ objects,~\cite{Doucot02,Rizzi06} so that the basic flux quantum
for a large closed chain of rhombi becomes $h/(4e)$ instead of $h/(2e)$
which can be directly observed by measuring the critical current of the loop
made from such chain and a large Josephson junction.

The main goal of the present paper is to discuss a practical way to probe
directly the quantum coherence associated with these tunneling processes
between macroscopically distinct classical ground-states. In principle, it
is relatively simple to implement, since it amounts to measuring the average
dc voltage generated across a finite Josephson junction array in the
presence of a small current bias (i. e. this bias current has to be smaller
than the critical current of the global system). The physical mechanism
leading to this small dissipation is very interesting by itself; it was
orinally discussed in a seminal paper by Likharev and Zorin \cite{Likharev85}
in the context of a single Josephson junction. Consider one element (single
junction or a rhombus) of the chain, and denote by $\phi $ the phase
difference across this element. When it is disconnected from the outside
world, its wave-function $\Psi $ is $2\pi \zeta $-periodic in $\phi $ where $%
\zeta =1$ for a single junction and $\zeta =1/2$ for a rhombus. This
reflects the quantization of the charge on the island between the elements
which can change by integer multiples of $2e/\zeta $. If $\phi $ is totally
classical, the element's energy is not sensitive to the choice of a
quasi-periodic boundary condition of the form 
\mbox{$\Psi(\phi+2\pi\zeta)=
\exp(i2\pi\zeta q)\Psi(\phi)$}, where $q$ represents the charge difference
induced across the rhombus. In the presence of coherent quantum tunneling
processes for $\phi $, the energy of the element $\epsilon (q)$ will acquire 
$q$-dependence, with a bandwidth directly related to the basic tunneling
amplitude. Whereas $q$ is constrained to be integer for an isolated system,
it is promoted to a genuine continuous degree of freedom when the array is
coupled to leads and therefore to a macroscopic dissipative environment. So,
as emphasized by Likharev and Zorin~\cite{Likharev85}, the situation becomes
perfectly analogous to the Bloch theory of a quantum particle in a
one-dimensional periodic potential, where the phase $\phi $ plays the role
of the position, and $q$ of the Bloch momentum. A finite bias current tilts
the periodic potential for the phase variable, so that in the absence of
dissipation, the dynamics of the phase exhibits Bloch oscillations, very
similar to those which have been predicted~\cite{Wannier62} and observed~%
\cite{Mendez88,Voisin88} for electrons in semi-conductor super-lattices. If
the driving current is not too large, it is legitimate to neglect inter-band
transitions induced by the driving field, and one obtains the usual spectrum
of equally spaced localized levels often called a Wannier-Stark ladder. In
the presence of dissipation, these Wannier-Stark levels acquire a finite
life-time, and therefore the time-evolution of the phase variable is
characterized by a slow and uniform drift superimposed on the faster Bloch
oscillations. This drift is translated into a finite dc voltage by the
Josephson relation \mbox{$2eV=\hbar(d\phi/dt)$}. This voltage
decreases with current until one reaches the current bias high enough to
induce the interband transition. At this point the phase starts to slide
down fast and the junction switches into a normal state. In the context of
Josephson junctions these effects were first observed in the experiments on
Josephson contacts with large charging energy\cite%
{Kuzmin1991,Kuzmin1994,Kuzmin1994b,Kuzmin1996} and more recently\cite%
{Watanabe03,Corlevi06} in the semiclassical (phase) regime of interest to us
here. Bloch oscillations in the quantronium circuit driven by a
time-dependent gate voltage have also been recently observed.~\cite%
{Boulant06}

This picture holds as long as the dissipation affecting the phase dynamics
is not too strong, so that the radiative width of the Wannier-Stark levels
is smaller than the nearest-level spacing (corresponding to phase
translation by $2\pi \zeta $) that is proportional to the bias current. This
provides a lower bound for the bias current which has to be compatible with
the upper bound coming from the condition of no inter-band transitions. As
we shall see, this requires a large real part of the external impedance $%
Z_{\omega }\gg R_{\mathrm{Q}}$ as seen by the element at the frequency of
the Bloch oscillation, where the quantum resistance scale is $R_{\mathrm{Q}%
}=h/(4e^{2})$. This condition is the most stringent in order to access
experimentally the phenomenon described here. Note that this physical
requirement is not limited to this particular experimental situation,
because any circuit exploiting the quantum coherence of phase variables, for
instance for quantum information processing, has to be imbedded in an
environment with a very large impedance in order to limit the additional
quantum fluctuations of the phase induced by the bath. The intrinsic
dissipation of Josephson elements will of course add to the dissipation
produced by external circuitry, but we expect that in the quantum regime
(i.e. with sizable phase fluctuations) considered here, this additional
impedance will be of order of $R_{\mathrm{Q}}$ at the superconducting
transition temperature, and will grow exponentially below. Thus, the success
of the proposed measurements is also a test of the quality of the
environment for the circuits intended to serve as protected qubits.

In many physical realizations $Z_{\omega }$ has a significant frequency
dependence and the condition $Z_{\omega }\gg R_{\mathrm{Q}}$ is satisfied
only in a finite frequency range $\omega _{\max }>\omega >\omega _{\min }$.
This situation is realized, for example, when the Josephson element is
decoupled from the environment by a long chain of larger Josephson junctions
(Section \ref{Implementations}). In this case the superconducting phase
fluctuations are suppressed at low frequencies implying that a phase
coherence and thus Josephson current reappears at these scales. The
magnitude of the critical current is however strongly suppressed by the
fluctuations at high frequencies. This behavior is reminiscent of the
reappearance of the Josephson coupling induced by the dissipative
environment observed in \cite{Steinbach2001}. At higher energy scales
fluctuations become relevant, the phase exhibits Bloch oscillations
resulting in the insulating behavior described above. Thus, in this setup
one expects a large hierarchy of scales: at very low currents one observes a
very small Josephson current, at larger currents an almost insulating
behavior and finally a switching into the normal state at largest currents.  

In the case of a chain of identical elements, the total dc voltage is
additive, but Bloch oscillation of different elements might happen either in
phase or in antiphase. In the former case the ac voltages add increasing the
dissipation in the external circuitry; while in the latter case the
dissipation is low and the individual elements get more decoupled from the
environment. As we show in Section \ref{sec-chain} a small intrinsic
dissipation of the individual elements is crucial to ensure the antiphase
scenario.

This paper is organized as follows. In section~\ref{Semiclassical}, we
present a semi-classical treatment of the voltage versus bias current curves
for a single Josephson element. We show that this gives an accurate way to
measure the effective dispersion relation $\epsilon (q)$ of this element,
which fully characterizes its quantum transition amplitude. Further, we show
that application of the ac voltage provides a direct probe of the
periodicity ($2\pi $ versus $\pi $) of each element. In Section \ref%
{sec-chain} we consider the chain of these elements and show that under
realistic assumptions about the dynamics of individual elements, it provides
much more efficient decoupling from the environment. Section~\ref%
{Energybands} focusses on the dispersion relation expected in a practically
important case of a fully frustrated rhombus which is the building block for
the protected arrays considered before.~\cite{Doucot03a,Doucot05} In this
case, the band structure has been determined by numerical diagonalizations
of the quantum Hamiltonian. An important result of this analysis is that
even in the presence of relatively large quantum fluctuations, the effective
band structure is always well approximated by a simple cosine expression.
Finally, in section~\ref{Implementations} we discuss the conditions for the
experimental implementation of this measurement procedure and the full $V(I)$
characteristics expected in realistic setup. After a Conclusion section, an
Appendix presents a full quantum mechanical derivation of the dc voltage
when the bias current is small enough so that inter-band transitions can be
neglected, and large enough so that the level decay rate can simply be
estimated from Fermi's golden rule.

\section{Semi-classical equations for a single Josephson element}

\label{Semiclassical}

\begin{figure}[h]
\includegraphics[width=2.0in]{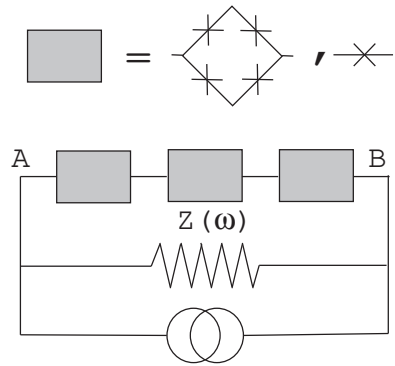}
\caption{The experimental setup discussed in this paper: a chain of
identical building blocks represented by shaded rectangle that are biased by
the external current source characterized by the impedance $Z(\protect\omega %
)$. The internal structure of the block that is considered in more detail in
the following sections is either a rhombus (4 junction loop)\ frustrated by
half flux quantum, or a single Josephson junction but the the results of the
section~\protect\ref{Semiclassical} can be applied to any circuit of this
form provided that the junctions in the elementary building blocks are in
the phase regime, i.e. $E_{J}\gg E_{C}$. }
\label{circuitry}
\end{figure}

Let us consider the system depicted on Fig.~(\ref{circuitry}). In the
absence of the current source, the energy of the one dimensional chain of $N$
Josephson elements is a $2\pi \zeta $ periodic function of the phase
difference $\phi $ across the chain. The current source is destroying this
periodicity by introducing the additional term $-\hbar (I/2e)\phi $ in the
system's Hamiltonian. Because $\phi $ is equal to the sum of phase
differences across all the individual elements, it seems that the voltage
generated by the chain is $N$ times the voltage of a chain reduced to a
single element. This is, however, not the case:\ the individual elements are
coupled by the common load, and furthermore, as we show in the next section,
their collective behavior is sensitive to the details of the single element
dynamics. In this section, we consider the case of a single Josephson
element ($N=1$), rederive the results of Likharev and Zorin~\cite{Likharev85}
for single Josephson contact and generalize them for more complicated
structures such as rhombus and give analytic equations convenient for data
comparison.

The dynamics of a single Josephson contact is analogous to the motion of a
quantum particle (with a charge $e$) in a one-dimensional periodic potential
(with period $a$) in the presence of a static and uniform force $F$, the
phase-difference $\phi $ playing the role of the spatial coordinate $x$ of
the particle.\cite{Likharev85} In the limit of a weak external force, it is
natural to start by computing the band structure $\epsilon _{n}(k)$ for $k$
in the first Brillouin zone $[-\pi /a,\pi /a]$, $n$ being the band label. A
first natural approximation is to neglect interband transitions induced by
the driving field. This is possible provided the Wannier-Stark energy gap %
\mbox{$\Delta_{\mathrm{B}}=Fa$} is smaller than the typical band gap $\Delta 
$ in zero external field. As long as $\Delta _{\mathrm{B}}$ is also smaller
than the typical bandwidth $W$, the stationary states of the Schr\"{o}dinger
equation spread over many (roughly $W/\Delta _{\mathrm{B}}$) periods, so we
may ignore the discretization (i.e. one quantum state per energy band per
spacial period) imposed by the projection onto a given band. We may
therefore construct wave-packets whose spacial extension $\Delta x$
satisfies \mbox{$a\ll \Delta x \ll a W/\Delta_{\mathrm{B}}$}, and the center
of such a wave-packet evolves according to the semi-classical equations: 
\begin{eqnarray}
\frac{dx}{dt} &=&\frac{1}{\hbar }\frac{d\epsilon _{n}(k)}{dk} \\
\frac{dk}{dt} &=&\frac{1}{\hbar }F
\end{eqnarray}%
In the presence of dissipation, the second equation is modified according
to: 
\begin{equation}
\frac{dk}{dt}=\frac{1}{\hbar }F-\frac{m^{\ast }}{\hbar \tau }\frac{dx}{dt}
\end{equation}%
where $m^{\ast }$ is the effective mass of the particle in the $n$-th band
and $\tau $ is the momentum relaxation time introduced by the dissipation.

In the context of a Josephson circuit, we have to diagonalize the
Hamiltonian describing the array as a function of the pseudo-charge $q$
associated with the $2\pi \zeta $ periodic phase variable $\phi $. The
quantity $q$ controls the periodic boundary condition imposed on $\phi $,
namely the system's wave-function is multiplied by $\exp (i2\pi q)$ when $%
\phi $ is increased by $2\pi \zeta $. From this phase-factor, we see that
the corresponding Brillouin zone for $q$ is the interval $[-1/2,1/2]$. For a
simple Josephson contact ($\zeta =1$), the fixed value of $q$ means that the
total number of Cooper pairs on the site carrying the phase $\phi $ is equal
to $q$ plus an arbitrary integer. For a doubly periodic element, such as
rhombus ($\zeta =1/2$), charge is counted in the units of $4e$. To simplify
the notations we assume usual $2\pi $ periodicity ($\zeta =1$) in this and
the following Sections and restore the $\zeta $-factors in Sections \ref%
{Energybands}, \ref{Implementations}. From the band structure $\epsilon
_{n}(q)$, we may write the semi-classical equations of motion in the
presence of the bias current $I$ and the outer impedance $Z$ as: 
\begin{eqnarray}
\frac{d\phi }{dt} &=&\frac{1}{\hbar }\frac{d\epsilon _{n}(q)}{dq}
\label{dphidt} \\
\frac{dq}{dt} &=&\frac{I}{2e}-\frac{Z_{\mathrm{Q}}}{Z}\frac{d\phi }{dt}
\label{dqdt}
\end{eqnarray}%
where we used the Josephson relation for the voltage drop $V$ across the
Josephson element as \mbox{$V=(\hbar/2e)(d\phi/dt)$} and defined $Z_{\mathrm{%
Q}}=\hbar /(4e^{2})$.

This semi-classical model exhibits two different regimes. Let us denote by $%
\omega _{\mathrm{max}}$ the maximum value of the \textquotedblleft group
velocity\textquotedblright\ $|d\epsilon _{n}(q)/(\hbar dq)|$. If the driving
current is small ($I<I_{\mathrm{c}}=2e\omega _{\mathrm{max}}Z_{\mathrm{Q}}/Z$%
), it is easy to see that after a short transient, the system reaches a
stationary state where $q$ is constant and: 
\begin{equation}
\frac{d\phi }{dt}=\frac{I}{2e}\frac{Z}{Z_{\mathrm{Q}}}
\end{equation}%
that is: $V=ZI$. Thus, at $I<I_{\mathrm{c}}$ the current flows entirely
through the external impedance, i.e. the Josephson elements become
effectively insulating due to quantum phase fluctuations. Indeed, a Bloch
state written in the phase reprentation corresponds to a fixed value of the
pseudo-charge $q$ and non-zero dc voltage $(1/2e)(d\epsilon _{n}/dq)$. Note
that the measurement of the maximal value $V_{\mathrm{c}}$ of the voltage on
this linear branch directly probes the spectrum of an individual Josephson
block, because $V_{\mathrm{c}}=\hbar \omega _{\mathrm{max}}/2e$

At stronger driving ($I>I_{\mathrm{c}}$), it is no longer possible to find a
stationary solution for $q$. The system enters therefore a regime of Bloch
oscillations. In the absence of dissipation ($Z/Z_{\mathrm{Q}}\rightarrow
\infty $), the motion is periodic in time for both $\phi $ and $q$. A small
but finite dissipation preserves the periodicity in $q$, but induces an
average drift in $\phi $ or equivalently a finite dc voltage. To see this,
we first note that the above equations of motion imply: 
\begin{equation}
\frac{dq}{dt}=\frac{I}{2e}-\frac{Z_{\mathrm{Q}}}{Z}\frac{1}{\hbar }\frac{%
d\epsilon _{n}}{dq}
\end{equation}%
Since the right-hand side is a periodic function of $q$ with period 1, $q(t)$
is periodic with the period $T(I)$ given by: 
\begin{equation}
T(I)=\int_{-1/2}^{1/2}f(q)dq
\end{equation}%
with 
\begin{equation}
f(q)=\left( \frac{I}{2e}-\frac{Z_{\mathrm{Q}}}{Z}\frac{1}{\hbar }\frac{%
d\epsilon _{n}}{dq}\right) ^{-1}
\end{equation}%
On the other hand, the instantaneous dissipated power reads: 
\begin{equation}
\frac{d}{dt}\left( \epsilon _{n}(q)-\frac{\hbar I}{2e}\phi \right) =-\hbar 
\frac{Z_{\mathrm{Q}}}{Z}(\frac{d\phi }{dt})^{2}
\end{equation}%
Because $q(t)$ is periodic, averaging this expression over one period gives: 
\begin{equation}
\langle \frac{d\phi }{dt}\rangle =\frac{2e}{I}\frac{Z_{\mathrm{Q}}}{Z}%
\langle (\frac{d\phi }{dt})^{2}\rangle
\end{equation}%
or, equivalently: 
\begin{equation}
\langle V\rangle =\frac{\hbar }{I}\func{Re}\left( \frac{Z_{\mathrm{Q}}}{Z}%
\right) \langle (\frac{d\phi }{dt})^{2}\rangle
\end{equation}%
Using the equations of motion, we get more explicitely: 
\begin{equation}
\langle V\rangle =\frac{1}{4e^{2}I}\func{Re}\left( \frac{Z_{\mathrm{Q}}}{%
Z_{\omega }}\right) \frac{\int_{-1/2}^{1/2}(\frac{d\epsilon _{n}}{dq}%
)^{2}f(q)dq}{\int_{-1/2}^{1/2}f(q)dq}
\end{equation}%
Here we emphasized by the subscript that $Z_{\omega }$ might have some
frequency dependence. As we show in Appendix, the dissipation actually
occurs at the frequency of Bloch oscillations that becomes $\omega _{\mathrm{%
B}}=2\pi I/2e$ in the limit of large currents. In the limit of large
currents, $I\gg I_{\mathrm{c}},$ (that can be achieved for large impedances)
we may approximate $f(q)$ by a constant, so the voltage is given by the
simpler expression: 
\begin{equation}
\langle V(I\gg I_{\mathrm{c}})\rangle =\frac{1}{4e^{2}Z_{\omega }I}%
\int_{-1/2}^{1/2}(\frac{d\epsilon _{n}}{dq})^{2}dq  \label{V_1}
\end{equation}%
On the other hand, when $I$ approaches $I_{\mathrm{c}}$ from above, Bloch
oscillations become very slow and $f(q)$ is strongly peaked in the vicinity
of the maximum of the group velocity. Since this velocity is in general a
smooth function of $q$, we get in this limit for the maximal dc voltage: 
\begin{equation}
V_{c}=\frac{\hbar ^{2}\omega _{\mathrm{max}}^{2}}{4e^{2}Z_{0}I_{\mathrm{c}}}%
=Z_{0}I_{\mathrm{c}}
\end{equation}

\begin{figure}[h]
\includegraphics[width=2.0in]{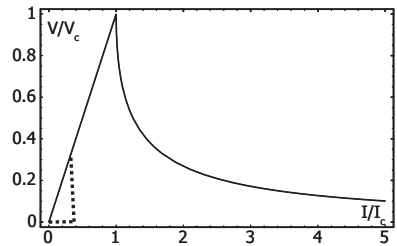}
\caption{Typical $I-V$ curve of a single Josephson element measured by a
circuits shown in Fig. 1}
\label{IVcurve}
\end{figure}

In the simplest case of a purely harmonic dispersion, $\epsilon (q)=2w\cos
2\pi q$, the maximal voltage $V_{c}=4\pi w/(2e).$ If one can further neglect
the frequency dependency of $Z$, the $V(I)$ can be computed analytically: 
\begin{eqnarray}
\left\langle V\right\rangle &=&ZI\;\;\;\;I<I_{\mathrm{c}}  \label{fullcurvea}
\\
\left\langle V\right\rangle &=&ZI_{\mathrm{c}}\frac{I_{\mathrm{c}}}{I+\sqrt{%
I^{2}-I_{\mathrm{c}}^{2}}}\;\;\;\;I>I_{\mathrm{c}}  \label{fullcurveb}
\end{eqnarray}
We show this dependence in Fig.~\ref{IVcurve}. This expression~(\ref%
{fullcurvea}), (\ref{fullcurveb}) is related to the known result for $Z\ll
Z_{\mathrm{Q}}$~\cite{Ivanchenko,Ingold} by the duality~\cite{Schmid}
transformation: 
\begin{eqnarray*}
V &\rightarrow &I, \\
I &\rightarrow &V, \\
Z &\rightarrow &\frac{1}{Z}.
\end{eqnarray*}

The semi-classical approximation is valid when the oscillation amplitude of
the superconducting phase is much larger than $2\pi $, which allows the
formation of the semi-classical wave-packets. When $I$ is much larger than $%
I_{\mathrm{c}}$, this oscillation amplitude is equal to $2eW/\hbar I$, where 
$W$ is the total band-width of $\epsilon _{n}(q)$. This condition also
ensures that the work done by the current source when the phase increases by 
$2\pi $ is much smaller than the band-width. In order to observe the region
of negative differential resistance, corresponding to the regime of Bloch
oscillations, we require therefore that: 
\begin{equation}
\frac{2\pi \hbar I_{\mathrm{c}}}{2e}\ll W\simeq \frac{2eV_{\mathrm{c}}}{\pi }%
,
\end{equation}%
where the last equality becomes exact in the case of a purely harmonic
dispersion. This translates into: 
\begin{equation}
Z\gg R_{\mathrm{Q}}.  \label{Zcondition}
\end{equation}

For large currents one can compute dc voltage directly by using the golden
rule (without semiclassics); we present the results in Appendix~\ref%
{Goldenrule}. The result is consistent with the large $I$ limit of Eq.~(\ref%
{fullcurveb}), $\left\langle V\right\rangle =V_{\mathrm{c}}^{2}/2ZI$. Deep
in the classical regime ($E_{J}\gg E_{C}$), the bandwidth and the generated
voltage become exponentially small. In this regime the bandwidth is much
smaller than the energy gaps, so these formulas are applicable (asuming (\ref%
{Zcondition}) is satisfied) until the splitting between Wannier-Stark levels
becomes equal to the first energy gap given by the Josephson plasma
frequency, i.e. for $I<e\omega _{J}/\pi $. Upon a further increase of the
driving current in this regime the generating voltage experiences resonant
increase for each splitting that is equal to the energy gap: $%
I_{k}=e(E_{k}-E_{0})/\pi .$ Physically, at these currents the phase slips
are rare events that lead to the excitation of the higher levels at a new
phase value that are followed by their fast relaxation. At very large
energies, the bandwidth of these levels becomes larger than their decay rate
due to relaxation, $(R_{\mathrm{Q}}/Z)E_{C}$. At these driving currents, the
system starts to generate large voltage and switches to a normal state. At a
very large $E_{J}$ this happens at the driving currents very close to the
Josephson critical current $2eE_{J}$, but in a numerically wide regime of $%
100\gtrsim E_{J}/E_{C}\gtrsim 10$ the generated voltage at low curents is
exponentially small but switching to the normal state occurs at
significantly smaller currents than $2eE_{J}$.

In the intermediate regime where $E_{J}$ and $E_{C}$ are comparable, we
expect a band-width comparable to energy gaps so that the range of
application of the quantum derivation is not much larger than the one for
the semi-classical approach.

Negative differential resistance associated to Bloch oscillations has been
predicted long ago,~\cite{Esaki70} and observed experimentally~\cite%
{Sibille90} in the context of semi-conductor superlattices. For Josephson
junctions in the cross-over regime ($E_{J}/E_{C}\simeq 1$), a negative
differential resistance has been observed in a very high impedance
environment,~\cite{Watanabe03} in good agreement with earlier theoretical
predictions.~\cite{Geigenmuller88} More recently, the $I-V$ curve of the
type shown on Fig~\ref{IVcurve} have been reported on a junction with a
ratio $E_{J}/E_{C}=4.5$~\cite{Corlevi06}. These experiments show good
agreement with a calculation which takes into account the noise due to
residual thermal fluctuations in the resistor.\cite{Beloborodov02}

Although the above results allows the extraction of the band structure of an
individual Josephson block from the measurement of dc $I-V$ curves, the
interpretation of actual data may be complicated by frequency dependence of
the external impedance $Z_{\omega }$. Additional information independent on $%
Z_{\omega }$ can be obtained from measuring the dc $V(I)$ characteristics in
the circuit driven by an additional ac current. In this situation, the
semi-classical equations of motion become: 
\begin{eqnarray}
\frac{d\phi }{dt} &=&\frac{1}{\hbar }\frac{d\epsilon _{n}(q)}{dq}
\label{dphidt2} \\
\frac{dq}{dt} &=&\frac{I+I^{\prime }\cos (\omega t)}{2e}-\frac{Z_{\mathrm{Q}}%
}{Z}\frac{d\phi }{dt}  \label{dqdt2}
\end{eqnarray}%
A small ac driving amplitude $I^{\prime }$ strongly affects the $V(I)$ curve
only in the vicinity of resonances where $n\omega _{\mathrm{B}%
}(I_{R})=m\omega $, with $m$ and $n$ integers. The largest deviation occurs
for $m=n=1$. Furthermore, for $I^{\prime }\ll I$ the terms with $m>1$ are
parametrically small in $I^{\prime }/I$ while for $I\gg I_{\mathrm{c}}$ the
terms with $n>1$ are parametrically small in $I_{\mathrm{c}}/I$.
Experimental determination of the resonance current, $I_{R}$, would allow a
direct measurement of the Bloch oscillation frequency and thus the
periodicity of the phase potential (see next Section). \textbf{\ }%
Observation of these mode locking properties have in fact provided the first
experimental evidence of Bloch oscillations in a single Josephson junction.~%
\cite{Kuzmin1991,Kuzmin1994}

We now calculate the shape of $V(I)$ curve in the vicinity of $m=n=1$ point
when both $I^{\prime}\ll I$ and $I \gg I_{\mathrm{c}}$. We denote by $%
\phi_0(t)$ and $q_0(t)$ the time-dependent solutions of the equations at $%
I=I_R$ in the absence of ac driving current. We shall look for solutions
which remain close to $\phi_0(t)$ and $q_0(t)$ at all times and expand them
in small deviations $\phi_1=\phi-\phi_0$, $q_1=q-q_0$. We can always assume
that $q_1$ has no Fourier component at zero frequency because such component
can be eliminated by a time translation applied to $q_0$. The equations for $%
\phi_1,q_1$ become

\begin{eqnarray}
\frac{d\phi_1 }{dt} &=&\frac{1}{\hbar } \epsilon^{\prime\prime}_{n}(q_0)q_1
\label{dphidt3} \\
\frac{dq_1}{dt} &=&\frac{I-I_R+I^{\prime}\cos(\omega t)}{2e}- \frac{Z_{%
\mathrm{Q}}}{Z}\frac{d\phi_1}{dt}  \label{dqdt3}
\end{eqnarray}

Because the main component of $\frac{d^{2}\epsilon _{n}(q_{0})}{dq^{2}}$
oscillates with frequency $\omega $ and $q_{1}$ has no dc component, the
average value of the voltage $\frac{d\phi _{1}}{dt}$ is due to the part of $%
q_{1}$ that oscillates with the same frequency, $q_{1\omega }=I^{\prime
}/(2e\omega )\sin (\omega t)$. Because $q_{0}=\omega (t-t_{0})+\chi (\omega
(t-t_{0}))$ where $\chi (t)$ is a small periodic function, the first
equation implies that 
\begin{eqnarray*}
<\frac{d\phi _{1}}{dt}> &=&\langle \frac{1}{\hbar }\epsilon _{n}^{\prime
\prime }(\omega (t-t_{0}))\sin (\omega t)\rangle \\
&&\mbox{}<\frac{1}{\hbar }\int_{0}^{1}\epsilon _{n}^{\prime \prime }(q)\cos
(2\pi q)dq
\end{eqnarray*}%
The deviation $q_{1}$ remains small only if the constant parts cancel each
other in the right hand side of the equation (\ref{dqdt3}). This implies 
\begin{equation}
<\frac{d\phi _{1}}{dt}>=\frac{Z}{Z_{\mathrm{Q}}}\frac{I-I_{R}}{2e}<\frac{1}{%
\hbar }\int_{0}^{1}\epsilon _{n}(q)\cos (2\pi q)dq
\end{equation}

We conclude that in the near vicinity of the resonances the increase of the
current does not lead to additional current through the Josephson circuit,
so the relation between current and voltage becomes linear again $\delta
V=Z\delta I$. In other words, the Josephson circuit becomes insulating with
respect to current increments. The width of this region (in voltage) is
directly related to the first moment of the energy spectrum of the Josephson
block providing one with the direct experimental probe of this quantity. In
particular, a Josephson element such as rhombus in a magnetic flux somewhat
different from $\Phi _{0}/2$ displays a phase periodicity $2\pi $ but a very
strong deviations from a simple $\cos 2\pi q$ spectrum that will manifest
themselves in first moment of the spectrum. Note finally, that the
discussion above assumes that the external impedance $Z_{\omega }$ has no
resonances in the important frequency range. The presence of such resonances
will modify significantly the observed $V(I)$ curves because it would
provide an efficient mechanism for the dissipation of Bloch (or Josephson)
oscillations at this frequency.

\section{Chain of Josephson elements}

\label{sec-chain}

We shall first consider the simplest example of a two-element chain, because
it captures the essential physics. This chain is characterized by two phase
differences ($\phi _{1}$ and $\phi _{2}$) and two pseudo-charges ($q_{1}$
and $q_{2}$). The equations of motion for the pseudo-charges~(\ref{dqdt})
implies that the charge difference $q_{1}-q_{2}$ is constant, because the
currents flowing through these elements are equal, and thus the right-hand
sides of the evolution equations~(\ref{dqdt}) are identical. Because of this
conservation law, even the long-term physical properties depend on the
initial conditions. Similar problems have already been discussed in the
context of a chain of Josephson junctions driven by a current larger than
the critical current.~\cite{Wiesenfeld89,Tsang91,Nichols92,Strogatz93} This
unphysical behavior disappears if we take into account the dissipation
associated with individual elements. Physically, it might be due to stray
charges, two-level systems, quasi-particles, phonon emission, etc.~\cite%
{Faoro06,Ioffe04} A convenient model for this dissipation is to consider an
additional resistor in parallel with each junction. For the sake of
simplicity, we assume that each element has a low energy band with a simple
cosine form. This physics is summarized by the equations: 
\begin{eqnarray}
\dot{\phi}_{j} &=&4\pi w\sin 2\pi q_{j} \\
\dot{q}_{j} &=&\frac{1}{2e}\left( I-\frac{1}{2eZ}\sum_{i}\dot{\phi}_i- \frac{%
1}{2eR_{j}}\dot{\phi}_{j}\right)
\end{eqnarray}%
Eliminating the phases gives: 
\begin{eqnarray*}
\left( \dot{q_{1}}+\Omega _{1}\sin 2\pi q_{1}\right) &=&\nu -\frac{\nu _{0}}{%
2}(\sin 2\pi q_{1}+\sin 2\pi q_{2}) \\
\left( \dot{q_{2}}+\Omega _{2}\sin 2\pi q_{2}\right) &=&\nu -\frac{\nu _{0}}{%
2}\left( \sin 2\pi q_{1}+\sin 2\pi q_{2}\right)
\end{eqnarray*}%
where 
\begin{eqnarray*}
\Omega _{i} &=&\frac{4\pi w}{(2e)^{2}R_{i}}, \\
\nu &=&\frac{I}{2e} \\
\nu _{0} &=&\frac{8\pi w}{(2e)^{2}Z}
\end{eqnarray*}%
Here we allowed for different effective resistances associated with each
element because this has an important effect on their dynamics. Indeed the
difference between the currents flowing through the resistors changes the
charge accumulated at the middle island and therefore violates the
conservation law mentioned before. Using the notations $\delta \Omega
=(\Omega _{2}-\Omega _{1})/2$ and $q_{\pm }=(q_{2}\pm q_{1})/2$, we have: 
\begin{eqnarray}
\dot{q_{-}}+\Omega \sin 2\pi q_{-}\cos 2\pi q_{+} + & &  \notag \\
+\delta \Omega \cos 2\pi q_{-}\sin 2\pi q_{+} &=&0  \label{q-} \\
\dot{q_{+}}+\left( \nu _{0}+\Omega \right) \sin 2\pi q_{+}\cos 2\pi q_{-}- &
&  \notag \\
-\delta \Omega \sin 2\pi q_{-}\cos 2\pi q_{+} &=&\nu  \label{q+}
\end{eqnarray}%
Significant quantum fluctuations imply that internal resistance of the
element $R\sim Z_{Q}$ for individual elements at $T\lesssim T_{C}$; at lower
temperature it grows exponentially. Thus, in a realistic case $R\gg Z$ which
implies that $\Omega _{i}\ll \nu $. In the insulating regime the equations (%
\ref{q-}-\ref{q+}) have stable stationary solution $\left( \nu _{0}+\Omega
\right) \sin 2\pi q_{+}=\nu $, $q_{-}=0$. This solution exists for $\left(
\nu _{0}+\Omega \right) <\nu $ , i.e. if the voltage drop across both
junctions does not exceed $V_{c}=8\pi w/(2e)$. The conducting regime occurs
when $\nu >\left( \nu _{0}+\Omega \right) $; to simplify the analytic
calculations we assume that $\nu \gg \nu _{0}$. This allows to solve the
equations (\ref{q-}-\ref{q+}) by iterations in all non-linear terms. In the
absence of non-linearity $q_{+}=\nu t$ , $q_{-}=const$; the first iteration
gives periodic corrections $\propto \cos 2\pi \nu t$. Averaging the result
of the second order iteration over the period we get 
\begin{equation}
\dot{\langle q_{-}\rangle }=-\frac{\delta \Omega }{2\nu }\left[ \nu _{0}\cos
^{2}2\pi q_{-}+2\Omega \right]  \label{q-dot}
\end{equation}%
The second term in the right hand side of this equation is much smaller than
the first if $\Omega \ll \nu _{0}$. In its absence the dynamics of $q_{-}$
has fixed points at $\cos 2\pi q_{-}=0$. At these fixed points the periodic
potentials generated by individual elements cancel each other and the
dissipation in external circuitry (which is proportional to $\cos ^{2}(2\pi
q_{-})$) is strictly zero. In a general case the equation (\ref{q-dot}) has
solution 
\begin{equation*}
\cos ^{2}(2\pi q_{-})=\frac{1}{1+\frac{\nu _{0}+2\Omega }{2\Omega }\tan ^{2}%
\left[ \frac{\pi }{\nu }\delta \Omega \sqrt{2\Omega (\nu _{0}+2\Omega )}t%
\right] }
\end{equation*}%
that corresponds to the short bursts of dissipation in external circuitry
that occur with low frequency $\nu _{b}=\frac{2}{\nu }\delta \Omega \sqrt{%
2\Omega (\nu _{0}+2\Omega )}$. The average value of $\cos ^{2}(2\pi q_{-})$ 
\begin{equation*}
<\cos ^{2}(2\pi q_{-})>=\frac{1}{1+\sqrt{\frac{\nu _{0}+2\Omega }{2\Omega }}}%
\approx \sqrt{\frac{2\Omega }{\nu _{0}+2\Omega }}
\end{equation*}%
is small implying that the effective dissipation introduced by the external
circuitry is strongly suppressed because the pseudocharge oscillations on
different elements almost cancel each other. The effective impedance of the
load seen by individual junction is strongly increased:%
\begin{equation}
Z_{eff}=\sqrt{\frac{\nu _{0}+2\Omega }{2\Omega }}Z  \label{Z_eff}
\end{equation}%
Similar to a single element case discussed in the previous Section, an
additional dissipation in the external circuit implies dc current across the
Josephson chain%
\begin{equation*}
V=V_{c}\frac{I_{c}}{2I}\;I\gg I_{c}=V_{c}/Z_{eff}
\end{equation*}

We conclude that a chain of Josephson elements has a current-voltage
characteristics similar to the one of the single element with one important
difference: the effective impedance of the external circuitry is strongly
enhanced by the antiphase locking of the individual Josephson elements. In
particular, it means that the condition $Z\gg R_{Q}$ is much easier to
satisfy for the chain of the elements than for a single element. The
analytical equations derived here describe the chain of two elements but it
seems likely that similar suppression of the dissipation should occur in
longer chains.

To substantiate this claim, lets us generalize the averaging method which
led to Eq.~(\ref{q-dot}) for $N=2$. The coupled equations of motion read: 
\begin{equation}
\dot{q_{j}}+\Omega _{j}\sin 2\pi q_{j}=\nu -\frac{\nu _{0}}{2}%
\sum_{k=1}^{N}\sin 2\pi q_{k}
\end{equation}%
To second order in $\Omega _{j}$ and $\nu _{0}$, the averaged equations of
motion are: 
\begin{eqnarray}
\left\langle \dot{q}_{j}\right\rangle &=&-\frac{\Omega _{j}^{2}}{2\nu }-%
\frac{\nu _{0}}{4\nu }\sum_{k=1}^{N}\Omega _{k}  \notag \\
&&\mbox{}-\frac{\nu _{0}^{2}}{8\nu }\sum_{k,l}\cos (2\pi (q_{k}-q_{l})) 
\notag \\
&&\mbox{}-\frac{\nu _{0}\Omega _{j}}{4\nu }\sum_{k=1}^{N}\cos (2\pi
(q_{j}-q_{k}))  \label{Naveraged}
\end{eqnarray}%
This set of coupled equations is similar to the Kuramoto model for coupled
rotors~\cite{Kuramoto84} defined as: 
\begin{equation}
\dot{q}_{j}=\omega _{j}-\frac{K}{N}\sum_{k=1}^{N}\sin (2\pi
(q_{j}-q_{k})+\alpha )  \label{Kuramoto}
\end{equation}%
The equation of motion (\ref{Kuramoto}) exhibits synchronisation of a finite
fraction of the rotors only for $K>K_{\mathrm{c}}(\alpha )$.~\cite%
{Sakaguchi86,Daido92} The last term in Eq.~(\ref{Naveraged}) is equivalent
to the interaction term of Kuramoto model \ with $\alpha =\pi /2$. The
additional (third) term in the model (\ref{Naveraged}) is the same for all
oscillators, it is thus not correlated with individual $q_{j}$\textbf{\ }and
thus can not directly lead to their synchronisation. Remarkably, it turns
out that for model (\ref{Kuramoto}) \mbox{$K_{\mathrm{c}}(\alpha=\pi/2)=0$}~%
\cite{Sakaguchi86,Daido92}, suggesting that in our case, synchronization
never occurs on a macroscopic scale. Note that the coupling $K$ arising from
Eq.~(\ref{Naveraged}) is not only $j$-dependent, but it is also proportional
to $N$. This could present a problem in the infinite $N$ limit, but should
not present a problem in a finite system. It is striking to see that $\alpha
=\pi /2$ is the value for which synchronization is the most difficult.

\section{Energy bands for a fully frustrated Josephson rhombus}

\label{Energybands}

In order to apply general results of the previous section to the physical
chains made of Josephson junctions or more complicated Josephson circuits we
need to compute the spectrum of these systems as a function of the
pseudocharge $q$ conjugated to the phase across these elements. In all cases
the superconducting phase in Josephson devices fluctuates weakly near some
classical value $\phi _{0}$ where the Josephson energy has a minimum in the
limit $E_{J}/E_{C}\gg 1$. In the vicinity of the minimum, the phase
Hamiltonian is $H=-4E_{C}\frac{d^{2}}{d\phi ^{2}}+\frac{1}{2}E^{\prime
\prime }(\phi _{0})(\phi -\phi _{0})^{2}$, so a higher energy state of the
individual element (at a fixed $q$) can be approximated by one of the
oscillator $E_{n}=(n+\frac{1}{2})\omega _{J}$ where the Josephson plasma
frequency $\omega _{J}=\sqrt{8E^{\prime \prime }(\phi _{0})E_{C}}\approx 
\sqrt{8E_{J}E_{C}}$. The Josephson energy is periodic in the phase with the
period $2\pi $ but the amplitude of the transitions between these minima is
exponentially small: 
\begin{equation*}
w=a\hbar \omega _{J}(E_{J}/E_{C})^{1/4}\exp (-c\sqrt{E_{J}/E_{C}})
\end{equation*}%
where $a,c\sim 1$. In this limit one can neglect the contribution of the
excited states (separated by a large gap $\omega _{J}$) to the lower band,
so the low energy spectrum acquires a simple form $\epsilon (q)=2w\cos 2\pi
q $. The numerical coefficients $c,a$ in the formulae for the transition
amplitude depend on the element construction. For a single junction $%
a_{s}=8\,2^{1/4}/\sqrt{\pi }$ , $c_{s}=\sqrt{8}$ while for the rhombus $%
a_{r}\approx 4.0$ , $c_{r}\approx 1.61$. In case of the rhombus in magnetic
field with flux $\Phi _{0}/2$ the Hamiltonian is periodic in phase with
period $\pi $ provided that the rhombus is symmetric along its horizontal
axis: indeed in this case the combination of the time reversal symmetry and
reflection ensures that the Josephson energy has a minimum for $\phi
_{0}=\pm \pi /2$. Thus, in this case the period in $q$ doubles and the low
energy band becomes $\epsilon (q)=2w\cos \pi q$. The maximal voltage
generated by the chain of $N$ such elements at $I=I_{c}=(8\pi \zeta ew/\hbar
)(Z_{\mathrm{Q}}/Z)$ is

\begin{equation}
V_{c}=N\frac{4\pi \zeta w}{2e}  \label{V_c_final}
\end{equation}

The voltage generated at larger currents depend on the collective behaviour
of the elements in the chain. For a single element it is simply

\begin{equation*}
\langle V(I)\rangle =\frac{(2\pi \zeta w)^{2}}{e^{2}Z_{\omega }^{{}}}\frac{1%
}{I+\sqrt{I^{2}-I_{\mathrm{c}}^{2}}},
\end{equation*}%
For more than one element the total volage is sufficiently reduced due to
the antiphase correlations. Generally, one expects that 
\begin{equation}
\langle V(I)\rangle =N\frac{(2\pi \zeta w)^{2}}{e^{2}Z_{\omega }^{eff}}\frac{%
1}{I+\sqrt{I^{2}-I_{\mathrm{c}}^{2}}},  \label{V_final}
\end{equation}%
where $Z_{\omega }^{eff}$ is the effective impedance of the environment
affecting each Josephson element which is generally much larger than its
'bare' impedance $Z_{\omega }^{{}}$. For two elements the exact solution
(see previous Section) gives $Z_{\omega }^{eff}\approx \sqrt{RZ}$ that shows
the increase of the effective impedance by a large factor $\sqrt{R/Z}$. We
expect that a similar enhancement factor appears for all $N\gtrsim 2$.
Finallly, For $I<I_{\mathrm{c}}$, the system is ohmic with: 
\begin{equation}
\langle V(I)\rangle =Z_{0}I
\end{equation}%
As discussed in Section \ref{Semiclassical}, application of a small
additional ac voltage produces features on the current-voltage
characteristics for the currents that produce Bloch oscillation with
frequencies commensurate with the frequency of the applied ac field $\omega
_{B}=2\pi \zeta I/2e=(m/n)\omega $. At these currents the system becomes
insulating with respect to current increments, the largest such feature
appears at $m=n=1$ that allows a direct measurement of the Josephson element
periodicity.

\begin{figure}[h]
\includegraphics[width=2.5in]{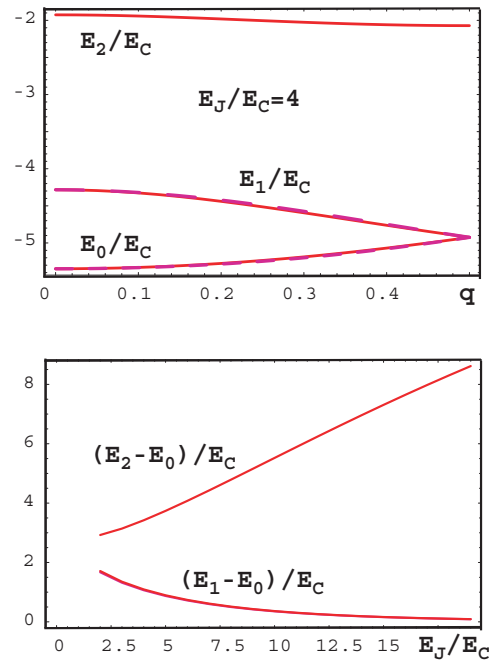}
\caption{ Spectrum of a single rhombus biased by magnetic flux $\Phi =\Phi
_{0}/2$. The upper pane shows the bands of the rhombus characterized by
Josephson eneergy $E_{J}/E_{C}=4$ as a function of bias charge, $q$. The two
lower levels are fitted by the first two harmonics (dashed line), the
coefficient $w^{\prime }$ of the second harmonics is $w^{\prime }=0.1w$. One
observes period doubling of the first two states that reflects the
symmetries of the rhombus frustrated by a half flux quantum. The second
excited level is doubly degenerate that makes its period doubling difficult
to observe. Physically, these states correspond to an excitation localized
on the upper or lower arms of the rhombus. The lower pane shows the
dependence of the gaps for $q=0$ as a function of $E_{J}/E_{C}$. Because
higher order harmonics are very small for all $E_{J}/E_{C}>1$, the gap $%
E_{1}-E_{0}$ coincides with $4w$ where $w$ is the tunneling amplitude
between the two classical ground states. }
\label{spectrum}
\end{figure}

For smaller $E_{J}/E_{C}\sim 1$ the quasiclassical formulas for the
transition amplitudes do not work and one has to perform the numerical
diagonalization of the quantum system in order to find its actual spectrum.
As $E_{J}/E_{C}\rightarrow 1$ the higher energy band approaches the low
energy band and the dispersion of the latter deviates from the simple cosine
form shown in Figure \ref{spectrum}. These deviations, lead to higher
harmonics in the dispersion: $\epsilon (q)=2w\cos 2\pi \zeta q+2w^{\prime
}\cos 4\pi \zeta q$ and change the equations (\ref{V_c_final},\ref{V_final}%
). Our numerical diagonalization of a single rhombus shows, however, that
even at relatively small $E_{J}/E_{C}\sim 1$ the second harmonics $w^{\prime
}$ does not exceed $0.15w$, so its additional contribution to the voltage
current characteristic ($\propto w^{\prime 2}$) can always be neglected.
Thus, in the whole range of $E_{J}/E_{C}>1$ the voltage current
characteristic is given by Eqs. (\ref{V_c_final},\ref{V_final}) where the
effective value of transition amplitude $t$ can be found from the band
width $W=E_{1}-E_{0}=4w$ plotted in Fig. \ref{spectrum}. For comparison we
show the variation of the lower band width for a single junction in Fig. \ref%
{singlejunctionspectrum}

\begin{figure}[h]
\includegraphics[width=2.5in]{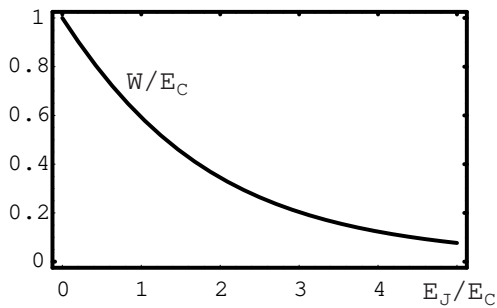}
\caption{Band width $W=4w$ of a single Josephson junction}
\label{singlejunctionspectrum}
\end{figure}

\section{Physical implementations}

\label{Implementations}

Generally, the effects described in the previous sections can be observed if
the environment does not affect much the quantum fluctuations of individual
elements and the resulting quasiclassical equations of motion. These
physical requirements translate into different conditions on the impedance
of the environment at different frequencies.

\bigskip We begin with the quantum dynamics of the individual elements.
The effect of the leads impedance on it can be taken into account by adding
the appropriate current term to the phase equation of motion before
projecting on a low energy band and requiring that their effect on the phase
dynamics is small at the relevant frequencies. For instance, for a single
junction 
\begin{equation*}
\frac{I}{2e}=E_{J}^{\prime }(\phi )+\frac{1}{4E_{c}}\frac{d^{2}\phi }{dt^{2}}%
+\left[ \frac{Z_{Q}}{Z_{\omega }}\right] \frac{d\phi }{dt}
\end{equation*}%
The characteristic frequency of the quantum fluctuations responsibe for the
tunneling of a single element is Josephson plasma frequency, $\omega _{J}=%
\sqrt{8E_{J}E_{c}}$, so the first condition implies that 
\begin{equation}
|Z(\omega _{J})|\gg \sqrt{E_{c}/E_{J}}Z_{Q}  \label{Z(omega_J)}
\end{equation}

For a typical $\omega _{J}/2\pi \sim 10GHz$, the impedance of a simple
superconducting lead of the length $\sim 1cm$ is smaller than $Z_{Q}$ and
the condition (\ref{Z(omega_J)}) is not satisfied. The situation is changed
if the Josephson elements are decoupled from the leads by a large resistance
or by a chain of $M\gg 1$ large junctions with $\sqrt{\widetilde{E}_{J}/%
\widetilde{E}_{c}}\gg 1$ that has no quantum tunneling transitions of their
own (the amplitude of such transitions is $\propto \exp (-\sqrt{8\widetilde{E%
}_{J}/\widetilde{E}_{c}}$ ). Assuming that elements of this chain have no
direct capacitive coupling to the ground ($M^{2}C_{0}\ll C$), the chain has
an impedance $Z=\sqrt{8\widetilde{E}_{c}/\widetilde{E}_{J}}MZ_{Q}$ at the
relevant frequencies, so a realistic chain with $M\sim 50$ junctions and $%
\sqrt{8\widetilde{E}_{J}/\widetilde{E}_{c}}\sim 10$ provides the
contribution to the impedance needed to satisfy (\ref{Z(omega_J)}).

Similar decoupling from the leads of the individual elements can be achieved
by a sufficiently long chain of similar Josephson elements, e.g. rhombi.
Consider a long ($N\gg 1$) chain of similar elements connected to the leads
characterized by a large but finite capacitance $C_{g}\gg C$. For a short
chain the tunneling of a single element changes the phase on the leads
resulting in a huge action of the tunneling process. However, in a long
chain of junctions, a tunneling of individual rhombus may be compensated by
a simultaneous change of the phases $\delta \phi /N$ of the remaining
rhombi, and subsequent relaxation of $\delta \phi $ from its initial value $%
\pi $ towards the equilibrium value which is zero. For $N\gg 1$, this later
process can be treated within the Gaussian approximation, with the
Lagrangian (in imaginary time): 
\begin{equation}
L=\frac{1}{16E_{g}}\dot{\phi}^{2}+\frac{1}{2N}E_{J}\phi ^{2}
\end{equation}%
where $E_{g}=e^{2}/(2C_{g})$. So the total action involved in the relaxation
is: \mbox{$S=\frac{\pi^{2}}{4\sqrt{2}}\left(\frac{E_J}{NE_g}\right)^{1/2}$}
If this action $S$ is less than unity this relaxation has strictly no effect
on the tunneling amplitude of the individual rhombus.

We now turn to the constraints imposed by the quasiclassical equations of
motion. The solution of these equations shows oscillation at the Bloch
frequency that is $\omega _{\mathrm{B}}=2\pi \zeta I/(2e)$ for large
currents and approaches zero near the $I_{c}$. Thus, for a single Josephson
element the quasiclassical equations of motion are valid if $\func{Re}%
(R_{Q}/Z(\omega _{\mathrm{B}}))\ll 1$ . A realistic energy band for a
Josephson element, $W\sim 0.3K$ and $Z/Z_{Q}\sim 100$ correspond to Bloch
frequency $\omega _{B}/2\pi \sim 0.1GHz$ . In this frequency range a typical
lead gives a capacitive contribution to the dynamics. The condition that it
does not affect significantly the equations of motion implies that the lead
capacitance $C\lesssim 10fF$. As discussed in Section (\ref{sec-chain}) the
individual elements in a short chain oscillate in antiphase decreasing the
effective coupling to the leads by a factor $\sqrt{R/Z}$ where $R$ is the
intrinsic resistance of the contact. This factor can easily reach $10^{2}$
at sufficiently low temperatures making much less restrictive the condition
on the lead capacitance.

Large but finite impedance of the environment $\func{Re}(R_{Q}/Z(\omega _{%
\mathrm{B}}))\lesssim 1$ modifes the observed current-voltage
characteristics qualitatively, specially in the limit of very small driving
current. When $I$ vanishes, and with infinite external impedance, the wave
function of the phase variable is completely extended, with the form of a
Bloch state, and the pseudo-charge $q$ is a good quantum number. As
discussed at the end of Sec.~\ref{Semiclassical}, the system behaves as a
capacitor. But when the external impedance is finite, charge fluctuations
appear, which in the dual description means that quantum phase fluctuations
are no longer unbounded. To be specific, consider a realistic example of $N$
rhombi chain (or two ordinary junctions) attached to the leads with $%
Z(\omega )=Z_{0}$ in a broad but finite frequency interval $\omega _{\min }<$
$\omega <$ $\omega _{\max }$ and decreases as $Z(\omega )=Z_{0}(\omega
_{\max }/\omega )$ for $\omega >$ $\omega _{\max }$, $Z(\omega
)=Z_{0}(\omega /\omega _{\min })$ for $\omega <$ $\omega _{\min }$. Such $%
Z(\omega )$ is realized in a long chain of $M$ Josephson junctions between
islands with a finite capactive coupling to the ground $C_{0}$: $\omega
_{\max }=\omega _{J}$ and $\omega _{\min }=(\sqrt{C/C_{0}}/M)\omega _{J}$.
The effective action describing the phase dynamics across the chain has
contributions from the tunneling of individual rhombi and from impedance of
the chain 
\begin{equation*}
L_{tot}=\frac{1}{2}\left[ \frac{\omega ^{2}}{8\pi ^{2}\zeta ^{2}Nw}+\frac{%
i\omega Z_{Q}}{Z(\omega )}\right] \phi ^{2}
\end{equation*}%
Here the first term describes the effect of the tunneling of the Josephson
element between its quasiclassical minima which we approximate by a single
tunneling amplitude $w$ resulting in a spectrum $\epsilon (q)=-2w\cos 2\pi
\zeta q$ that in a Gaussian approximation becomes $\epsilon (q)=4\pi
^{2}\zeta ^{2}wq^{2}$. This approximation is justified by the fact that, as
we show below, the main effect of the phase fluctuations comes from the
broad frequency range where the action is dominated by the second term while
the first serves only as a cutoff of the logarithmical divergence. Its
precise form is therefore largely irrelevant.

This action leads to a large but finite phase fluctuations%
\begin{equation*}
\left\langle \phi ^{2}\right\rangle =i\int \frac{d\omega }{2\pi }\frac{1}{%
\frac{\omega ^{2}}{8\pi ^{2}\zeta ^{2}Nw}+\frac{i\omega Z_{Q}}{Z(\omega )}}%
\approx \frac{Z_{0}}{R_{Q}}\ln \frac{\min (\omega _{\max },\omega _{\max
}^{\prime })}{\omega _{\min }}
\end{equation*}%
where $\omega _{\max }^{\prime }=8\pi ^{2}\zeta ^{2}Nw(Z_{Q}/Z_{0})$. These
fluctuations are only logarithmically large, so they result in a finite
renormalization of the Josephson energy of the rhombi chain and the
corresponding critical current. In the absence of such renormalization the
Josephson energy of a finite chain of elements can be approximated by the
leading harmonics $E(\phi )=-E_{0}\cos (\phi /\zeta )$ with $E_{0}\sim E_{J}$
for $N\sim 1$ and $E_{J}\gtrsim E_{c}.$ Renormalization by fluctuations
replaces $E_{0}$ by 
\begin{equation*}
E_{R}=\exp (-\frac{1}{2}\left\langle \phi ^{2}\right\rangle )E_{0}=\left[ 
\frac{\min (\omega _{\max },\omega _{\max }^{\prime })}{\omega _{\min }}%
\right] ^{-\frac{Z_{0}}{2R_{Q}}}E_{0}
\end{equation*}

In the limit of $\omega _{\min }\rightarrow 0$ or $Z_{0}\rightarrow \infty $
the phase fluctuations renormalize Josephson energy to zero. But for
realistic parameters this suppression of Josephson energy is finite which
thus results in a small but non-zero value of the critical current. In this
situation the current-voltage characterictics sketched in Fig.~\ref%
{circuitry} is modified for very small values of currents and voltages:
instead of insulating regime at very low currents and voltages one should
observe a very small supercurrent ($E_{R}/2e$) followed by a small voltage
step as shown in Fig. \ref{IVcurve} by a \ dashed line. As is clear from the
above discussion the value of the resulting critical current is controlled
by the phase fluctuations at low $\omega \ll \omega _{\max }$; these
frequencies are much smaller than the typical internal frequencies of a
chain of Josephson elements which can be thus lumped together into an
effective object characterized by the bare Josephson energy $E(\phi )$ and
transition amplitude between its minima $w$. We thus expect the same
qualitative behavior for a small chain of Josephson elements as for a single
element at low currents.

\section{Conclusion}

The main results of the present work are the expressions~(\ref{V_c_final}), (%
\ref{V_final}) for the I-V curves of a chain of $N$ identical basic
Josephson circuits. They are derived within the assumption that the
Josephson coupling is much larger than the charging energy, but in fact, the
numerical calculations show that they remain very accurate even if $%
E_{J}\approx E_{C}$. These equations predict a maximum dc voltage when $I=I_{%
\mathrm{c}}$ and $V(I)\propto 1/I$ for $I\gg I_{\mathrm{c}}$. The anomalous $%
V$ versus $I$ dependence exhibited by these equations is a signature of
underdamped quantum phase dynamics. It occurs only if the impedance of the
external circuitry is sufficiently large both at the frequency of Bloch
oscillations and at the Josephson frequency of the individual elements. The
precise conditions are given in Section \ref{Implementations}. Observation
of this dependence and the measurement of the maximal voltage would provide
the proof of the quantum dynamics and the measurements of the tunneling
amplitude which is the most important characteristics of these systems. It
would also provide a crucial test of the quality of decoupling to the
environment.

As a deeply quantum mechanical system, the chain of Josephson devices is
very sensitive to an additional ac driving. It exhibits resonances when the
driving frequency is commensurate with the frequency $\omega _{B}=2\pi \zeta
I/2e$ of the Bloch oscillations. This would provide additional ways to
characterize the quantum dynamics of these circuits and confirm the period
doubling of the rhombi frustrated by exactly half flux quantum.

\textbf{Acknowledgments}

LI is thankful to LPTMS Orsay, and LPTHE Jussieu for their hospitality
through a financial support from CNRS while BD has enjoyed the hospitality
of the Physics Department at Rutgers University. This work was made possible
by support from NSF DMR-0210575, ECS-0608842 and ARO W911NF-06-1-0208.

\appendix

\section{Quantum-mechanical calculation of the dc voltage}

\label{Goldenrule}

In the large current regime where $I \geq I_{\mathrm{max}} $, the energy
drop $\Delta_{\mathrm{B}}=hI/2e$ induced by the driving current when $\phi$
increases by $2\pi$ becomes comparable to or larger than the bandwidth $W$.
In this regime, the semi-classical approach is no longer reliable. But as
long as $\Delta_{\mathrm{B}}$ remains small compared to the typical gap $%
\Delta$ between nearby bands, we may still construct the system wave
functions in the presence of the driving field from Wannier orbitals
belonging to a single band. In such quantum-mechanical approach, dissipation
is described as the result of coupling the single degree of freedom $%
(\phi,q) $ to a continuum of oscillator modes $(q_{\alpha},p_{\alpha})$. The
corresponding Hamiltonian has the form: 
\begin{equation}  \label{HA1}
H_{n}=\epsilon_{n}\left(q-\sum_{\alpha}g_{\alpha}q_{\alpha}\right) -\frac{%
\hbar I}{2e}\phi+ \sum_{\alpha}\frac{\hbar\omega_{\alpha}}{2}%
(q_{\alpha}^{2}+p_{\alpha}^{2})
\end{equation}
where we have chosen the following commutation relations: 
\begin{equation}
[\phi,q] = i,\;\;\;\; [q_{\alpha},p_{\beta}] = i \delta_{\alpha\beta}
\end{equation}
and all other commutators between these operators vanish. The form of~(\ref%
{HA1}) is plausible on the physical ground because when the superconducting
islands are coupled to macroscopic leads, the charge $q$ undergoes quantum
fluctuations, so that it has to be replaced by a ``dressed charge'' %
\mbox{$q-\sum_{\alpha}g_{\alpha}q_{\alpha}$} in the effective Hamiltonian. A
more explicit justification is that the corresponding semi-classical
equations of motion have the same form as Eqs.~(\ref{dphidt}),~(\ref{dqdt})
which simply mean that the effective current going through the
superconducting circuit is the bias current minus the current going through
the external impedence. The semi-classical equations deduced from~(\ref{HA1}%
) read: 
\begin{eqnarray*}
\frac{d\phi}{dt} & = & \frac{1}{\hbar}\frac{d\epsilon_{n}}{dq}
(q-\sum_{\alpha}g_{\alpha}q_{\alpha}) \\
\frac{dq}{dt} & = & \frac{I}{2e} \\
\frac{dq_{\alpha}}{dt} & = & \omega_{\alpha}p_{\alpha} \\
\frac{dp_{\alpha}}{dt} & = & -\omega_{\alpha}q_{\alpha} +\frac{g_{\alpha}}{%
\hbar}\frac{d\epsilon_{n}}{dq} (q-\sum_{\alpha}g_{\alpha}q_{\alpha})
\end{eqnarray*}
It is then natural to introduce \mbox{$q'=q-\sum_{\alpha}g_{\alpha}q_{%
\alpha}$}, so that: 
\begin{eqnarray}
\frac{d\phi}{dt} & = & \frac{1}{\hbar}\frac{d\epsilon_{n}}{dq}(q^{\prime}) \\
\frac{dq^{\prime}}{dt} & = & \frac{I}{2e}-\sum_{\alpha}g_{\alpha}\frac{%
dq_{\alpha}}{dt}  \label{dq'dt}
\end{eqnarray}
To show that~(\ref{dq'dt}) has the same form as~(\ref{dqdt}), we notice that
the driving term for the bath oscillators is directly proportional to $%
d\phi/dt$. Specifically: 
\begin{equation}
\frac{dq_{\alpha}}{dt}+\omega_{\alpha}^{2}q_{\alpha}=
\omega_{\alpha}g_{\alpha}\frac{d\phi}{dt}
\end{equation}
Going to Fourier space, we see that after averaging over initial conditions
for the bath oscillators, Eq.~(\ref{dq'dt}) takes the form: 
\begin{equation}
-i\omega\tilde{q}^{\prime}(\omega)=\frac{I}{2e}2\pi\delta(\omega)
-i\sum_{\alpha}g_{\alpha}^{2}\frac{\omega\omega_{\alpha}} {%
\omega^{2}-\omega_{\alpha}^{2}}\left(-i\omega\tilde{\phi}(\omega) \right)
\end{equation}
This is exactly the frequency space version of Eq.~(\ref{dqdt}), where as
usual, the dissipation is related to the spectral density of the bath by: 
\begin{equation}
i\sum_{\alpha}g_{\alpha}^{2}\frac{\omega\omega_{\alpha}} {%
\omega^{2}-\omega_{\alpha}^{2}}= \frac{Z_{\mathrm{Q}}}{Z(\omega)}
\end{equation}
Here we emphasize that $Z$ is typically frequency dependent, in which case
the term $(Z_{\mathrm{Q}}/Z)(d\phi/dt)$ in Eq.~(\ref{dqdt}) becomes a
convolution with $Z_{\mathrm{Q}}/Z$ replaced by a non-local kernel in time.

Now we turn to the solution of the quantum problem~(\ref{HA1}) in the large
driving current regime. Let us first consider the Josephson array without
dissipation. It is straightforward to express its eigenstates in the $q$
representation~\cite{Anderson63} because the Schr\"odinger equation reads
then: 
\begin{equation}
E\psi(q)=\epsilon_{n}(q)\psi(q)-i\frac{\hbar I}{2e}\frac{d\psi}{dq}(q)
\end{equation}
so that: 
\begin{equation}
\psi(q)=\psi(0)\exp\left(-i\frac{2e}{\hbar I}\int_{0}^{q}(\epsilon_{n}(q^{%
\prime})-E)dq^{\prime}\right)
\end{equation}
The energy spectrum is determined via the boundary condition $%
\psi(q+1)=\psi(q)$ so that: 
\begin{equation}
E_{\nu}=\int_{-1/2}^{1/2}\epsilon_{n}(q^{\prime})dq^{\prime}+\Delta_{\mathrm{%
WS}}\nu,\;\;\;\;\nu\;\mathrm{integer}
\end{equation}
This yields a Wannier-Stark ladder of spacially localized states, with a
constant level spacing equal to $\Delta_{\mathrm{B}}$. Note that increasing $%
\nu$ by one unit multiplies the wave-function $\psi(q)$ by $\exp(i2\pi q)$.
In the phase representation, this is equivalent to a translation by $-2\pi$.
Of course, in the absence of dissipation, these levels have infinite
life-time, and therefore, no dc voltage is generated.

Let us now consider the limit of a weak coupling to the dissipative bath.
This means that the decay rate $\Gamma$ of the Wannier-Stark levels is much
smaller than the level spacing $\Delta_{\mathrm{B}}$. Assuming that
transitions take place mostly between two adjacent levels, we get an average
voltage: 
\begin{equation}
\langle V \rangle = \frac{\hbar}{2e}\langle \frac{d\phi}{dt} \rangle = \frac{%
h\Gamma}{2e}
\end{equation}
The rate $\Gamma$ is estimated via Fermi's golden rule which we prefer to
use in the correlation function formulation : 
\begin{equation}
\Gamma_{\nu \rightarrow \nu^{\prime}}=\frac{1}{\hbar^{2}} |\langle \nu|\frac{%
d\epsilon_{n}}{dq}|\nu^{\prime}\rangle|^{2}\tilde{C}_{AA}
((\nu-\nu^{\prime})\omega_{\mathrm{B}} )
\end{equation}
where $\omega_{\mathrm{B}} =\Delta_{\mathrm{B}} /\hbar=2\pi I/(2e)$. In this
expression, $\tilde{C}_{AA}$ is the Fourier transform of the correlation
function, $C_{AA}(t-t^{\prime})=\langle A(t)A(t^{\prime})\rangle$ of the
Heisenberg operators $A(t)=\sum_{\alpha} g_{\alpha}q_{\alpha}(t)$, taken in
the equilibrium state of the dissipative bath.

We evaluate now the matrix element of the velocity operator $d\epsilon
_{n}/dq$ between Wannier-Stark states: 
\begin{equation}
\langle \nu |\frac{d\epsilon _{n}}{dq}|\nu ^{\prime }\rangle
=\int_{-1/2}^{1/2}\frac{d\epsilon _{n}}{dq}(q)\exp (i2\pi (\nu ^{\prime
}-\nu )q)\:dq
\end{equation}%
As we have seen, in most physically interesting situations, we can
approximate the periodic function $\epsilon (q)$ by a single harmonic $%
2w\cos (2\pi q)$. In this case: 
\begin{equation}
\langle \nu |\frac{d\epsilon _{n}}{dq}|\nu ^{\prime }\rangle =2\pi w\delta
_{|\nu ^{\prime }-\nu |,1}
\end{equation}%
In the zero temperature limit, the bath correlation function is: 
\begin{equation}
\tilde{C}_{AA}(\omega )=\sum_{\alpha }\pi g_{\alpha }^{2}\delta (\omega
-\omega _{\alpha })=\frac{2}{\omega }\func{Re}\left( \frac{Z}{Z_{\omega }}%
\right) \theta (\omega )
\end{equation}%
where $\theta (\omega )$ is the Heaviside step function. Putting all these
elements together gives: 
\begin{equation}
\langle V(I\gg I_{\mathrm{c}})\rangle =\frac{(2\pi w)^{2}}{2e^{2}Z_{\omega }I%
}
\end{equation}%
where $Z_{\omega }$ denotes the external impedance taken for %
\mbox{$\omega=\omegaWS$}, and this result is in perfect agreement with the
large current limit of the semi-classical treatment, shown as Eq.~(\ref%
{V_final}).

When the band structure is replaced by $2w\cos (2\pi \zeta q)$ as in the
case of a rhombus (for which $\zeta =1/2$), we have now: 
\begin{equation}
\langle V\rangle =\zeta \frac{h\Gamma }{2e}
\end{equation}%
The frequency of Bloch oscillations becomes $\omega _{\mathrm{B}}(\zeta
)=\zeta \omega _{\mathrm{B}}(\zeta =1)$, and the matrix element is
multiplied by $\zeta $, so that the voltage is multiplied by $\zeta ^{2}$.
Again, this is compatible with the semi-classical result~(\ref{V_final}).

\end{document}